\date{}
\documentstyle[aps,multicol,prl,epsf]{revtex}
\begin{document}
\draft

\title
{Persistent currents in diffusive metallic cavities: Large values 
and anomalous scaling with disorder }
\author{G. Chiappe and  M. J. S\'anchez}
\maketitle
\noindent
\begin{center}
\center{\it Departamento de F\'{\i}sica  J. J. Giambiagi,\\
Facultad de Ciencias Exactas  y Naturales,  \\
Universidad de Buenos Aires.
Ciudad Universitaria, 1428 Buenos Aires, Argentina.}\\ 
\end{center}
\begin{abstract}

The effect of disorder  on confined metallic cavities  with an Aharonov-Bohm 
flux line is addressed. 
We find that, even deep in the diffusive regime, large values of 
persistent currents may arise for a wide variety of geometries. 
We present numerical results  supporting  an anomalous scaling 
law  of the average typical current $< I_{typ}>$ with the strength
of disorder $w$, $< I_{typ}> \sim  w^{- \gamma}$ with $\gamma < 2$. 
This is contrasted with previously reported results 
obtained for cylindrical samples where a scaling   $< I_{typ}> \sim  w^{-2}$ 
has been found.
Possible links to, up to date, unexplained experimental data are finally 
discussed.  
\end{abstract}

\begin{multicols}{2}

Modern   advances in  nanotechnology  made possible  to manufacture systems 
in the submicron range.
For those  structures and at  temperatures of few $mK$, the quantum 
mechanical coherence of the electronic wave functions dominates the transport
of charge \cite{kouwen}. 

Recently, billiards with rough boundaries  have been considered in order to 
analyze the effect of generic boundary deformations on the spectral and 
eigenfunction properties of  ballistic samples \cite{cuevas,frahm,blan}. 
In addition, during  the  last  years, the magnetic response 
and the persistent currents of integrable 
and   chaotic  billiards  in the presence of a 
magnetic flux have been the object of many research papers 
\cite{vonoppen,klaus}.
Most  of them were inspired in the experiments of 
Ref.\cite{levy,mailly} in which the magnetic susceptibility and the 
persistent currents of mesoscopic semiconductor samples  were measured. 
Continuous billiard systems posses an  energy dispersion relation quadratic 
in the wave number $k$ and  are reasonable models  to describe  semiconductor 
quantum dots in the ballistic regime.
The situation is qualitative different for metallic systems. 
The breaking of the continuous translational symmetry by the lattice leads to 
a band structure and to a dispersion relation in $k$ which is
equivalent to that of the free electron model only in the large wavelength 
limit. 
Moreover, the {\sl metallic regime} is characterized by a half filled band and
a  typical electronic wavelength of the order of the lattice constant.
Also and unlike high mobility heterostructures, a realistic
model for a metallic sample should contemplate the effect of the elastic 
scattering with random impurities.

The observed persistent currents (PC) in metallic samples 
\cite{chandra,mohanty} have been  typically much larger than the 
predicted theoretical estimates that take into account the effect of 
static impurities and  much effort is still devoted to understand 
the nature of the discrepancy between theory and experiments 
\cite{guhr}. 
In a recent work we have shown that by effect of confinement,
large values of PC can arise in the half filling regime  for clean 
metallic cavities of different  shapes  \cite{acs}. 

In this letter  we  address  the effect of bulk disorder in   
confined metallic systems threaded by a magnetic flux. 
We are mainly concerned with  the interplay between confinement and disorder 
effects and in particular  with the  sensitivity  of  the persistent current 
to the   strength of disorder $w$. Our main result is  that  even deep in 
the diffusive regime large values of persistent currents can still be obtained 
for a wide variety of sample shapes. 

We employ the spinless tight-binding hamiltonian defined on a square lattice
and introduced in Ref.\cite{acs}, with the additional inclusion of diagonal 
disorder.  Once the profile of the sample is defined, the on-site energies 
are taken randomly distributed 
in the interval $[- w/2,w/2]$. We  consider non-single connected loops of 
various shapes with  up to 3 000 sites and  values of   $w$ ranging  from 
0.02 to 5 in order to study the transition between the ballistic and the 
diffusive  transport  regimes. In the following all the lengths will be 
in units of the lattice parameter $a$, and the magnetic flux $\alpha$ will be 
measured in units of $\phi_{o}= h c /e$.
An isolated non-single connected sample threaded by a 
magnetic flux  $\alpha$ carries a persistent current  $I$, even in the case 
of multiple scattering by static impurities. 
 For a sample containing  $N$ non-interacting electrons, the total current
can be calculated at zero temperature as 
$I \sim - {\partial E /\partial \alpha} \; $ with $E$ the total energy of the 
system \cite{cheung2}.
In the absence of disorder, the typical current  
$I_{typ} \equiv \sqrt{\int_{0}^{1} I^2 \; d\alpha}$ gives a measure of the 
magnitude of the current and  is a (fluctuating) function of $N$.
We denote its value at half filling  $I_{o}$.

For a finite value of  $w$, we define the average typical current 
$< I_{typ}> \equiv \sqrt {<I_{typ}^2>_{\delta n}}$ performing 
averages in a small 
energy window containing $1<< \delta n << N$ levels. 
It was previously checked that 
this procedure is equivalent to average at the 
same time on the disorder configuration ($w$) and on the number of levels 
($\delta n$) that is, $<...>_{w, \delta n} = <...>_{\delta n}$ \cite{mbsf}. 

\begin{figure}
\epsfysize=4in
\centerline{\epsffile{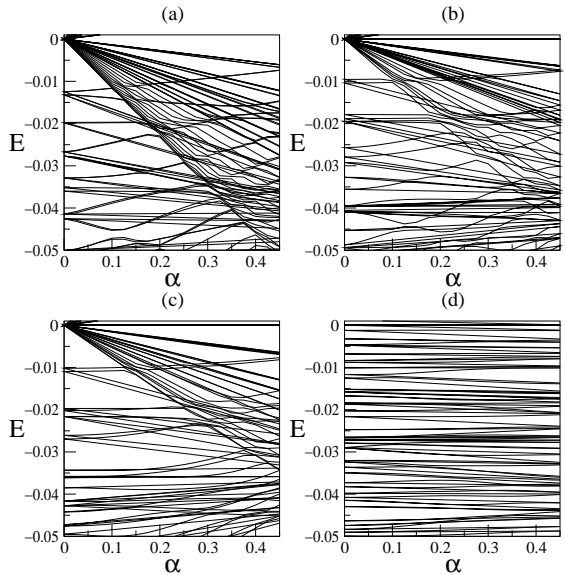}}
\epsfxsize=4in
\narrowtext
\caption{Energy levels   close to the band center as a function of 
$\alpha$ for: (a) SL in which the external (internal) square has 
a side $L_1=53 (L_2=13)$, (b) an AL loop with external (internal) radius 
$R=30 (r=3)$; (c) an  OL loop with the inner (outer) octagon inscribed in a 
circumference of $r=3 (R= 32)$ and (d) a PL with the inner (outer) pentagon 
inscribed in a circumference of radius  $r= 3(R= 35 $ ).}\label{1}
\end{figure}

Before considering disorder effects, we summarize the main results obtained 
for clean geometries.
In Ref.\cite{acs} we show that for annular and square loops the corresponding 
density of states (DOS) has a local peak close to the band  center 
$(E=0)$ at zero magnetic flux.
In the absence of surface roughness (SR), the peak consists of
border states that have in
general large values of mean angular momentum. These states contribute  
to enhance  in one or two orders of magnitude the value of 
$I_{o}$ with respect to values obtained for 
geometries with the same number of sites but without border states.
The peak in the DOS is robust to changes in  
the size of the sample while preserving  its shape.
In addition, we here present  numerical evidence supporting the conjecture 
that the peak in the DOS is present when the symmetry of the underlying 
lattice conforms a symmetry subgroup of the confining potential.
This  is illustrated in Fig.~\ref{1} , where 
 the spectra for four different samples are shown   close to half filling: (a) 
square SL, (b) annular AL, (c) octagonal OL and (d) pentagonal PL loops 
\cite{note}. In  Figs.~\ref{1}(a), (b) and (c), a bunch of degenerate levels 
is observed at zero flux for $E=0$. These levels  evolve paramagnetically  
for finite $\alpha$, generating  large values of PC .
Due to absence of crossings between {\rm occupied} and {\rm empty} levels 
close to half filling, $I_{typ}$ increases monotonically with the number of 
particles \cite{acs}. 
The PL, Fig.\ref{1}(d), corresponds to a generic geometry
where the mentioned  symmetry condition is not fulfilled. There is no
bunch of states at zero  flux, and therefore no enhancement of
the  PC near half filling is expected.

\begin{figure}
\epsfysize=4in
\centerline{\epsffile{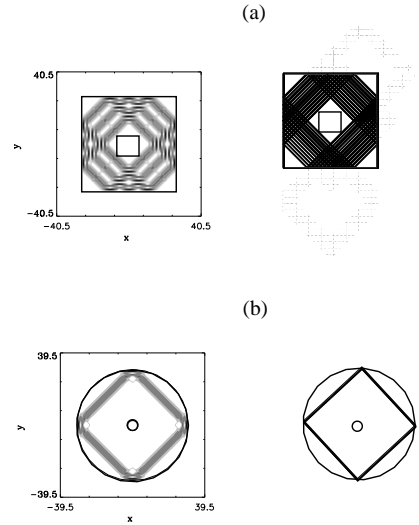}}
\epsfxsize=4in
\narrowtext
\vspace{-.1cm}
\caption{(a) Contour plot of the  absolute value square of the 
wave function for a  border state of  the SL for  
$\alpha = 0.025 $ (left panel). Sketch of the 
family of classical periodic orbits that support the quantum state (right 
panel). (b) diamond-like border state for the AL (left panel) 
togheter with  the associated classical orbit (right panel). See the text for
more details.}
\label{2} 
\end{figure}

When the confining potential has not  the square or rectangular rotational
symmetries  our  samples posses  surface roughness (SR). 
This is the case for the AL  and the OL in which, besides the 
border states that carry large currents, we find localized states  
in the bunch of degenerate levels at zero flux. These are states  
trapped by the SR and are not sensitive to the external 
magnetic flux. The flat lines on the top of  the spectra of Figs.~\ref{1} (b)
 and (c) correspond to these states.    
A remarkable feature is that the  charge distribution   for the border 
states seems to be  highly concentrated  on classical
 periodic trajectories that bounce off the  borders  of the 
loops. We find, as in  quantum billiard models, border states that
look like   {\rm resonant} states associated  to  a continuous set of  non 
isolated periodic orbits \cite{narevich2}. 
In the left panel of Fig.\ref{2} (a)  we show for the square loop SL, 
the contour plot of the  absolute  squared value of the wave
functions $(|\Psi(x,y)|^2)$ for the $20^{th}$ state  below  half 
filling  for $\alpha = 0.025$. In the right panel  of
Fig.~\ref{2} (a) the family of classical orbits that supports the 
quantum state is drawn for comparison. 
Fig.\ref{2} (b) shows  for the  annular loop AL the contour plot  for
a quantum state with  energy close to zero (that is belonging to the 
peak in the DOS) and  with the charge distribution extended along a 
diamond-like orbit that goes around the sample with large  mean angular 
momentum. As a consequence of the SR there is not a complete family of  
trajectories that supports the quantum state like in the square loop.

In summary  we have found, as a consequence of the symmetry
properties of the confining potential and the underlying lattice, clean 
metallic geometries with large values of PC at half filling.
We name these kind of samples {\it symmetric } loops.  
However, a realistic model for a metallic cavity should contemplate the 
inclusion of bulk disorder in order to mimic  the diffusive transport regime. 

\begin{figure}
\epsfysize=4in
\centerline{\epsffile{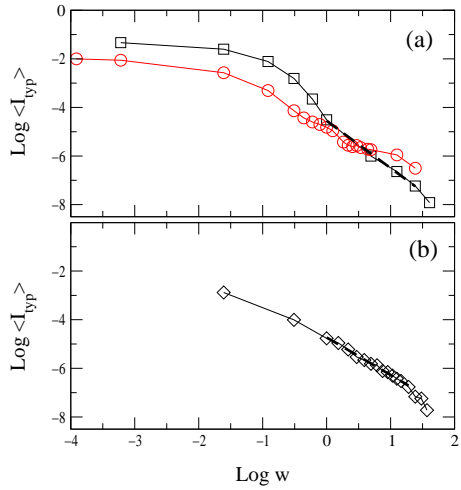}}
\epsfxsize=4in
\narrowtext
\vspace{-.5cm}
\caption{(a): Log-Log plot of  $<I_{typ}>$ as a function of $w$ for the 
AL  (circles) and for the SL (squares). The dashed 
line shows  the best fit $<I_{typ}> \sim w^{-1.98}$ for the SL. (b): Idem but
for a non-symmetric (chaotic) rectangular loop  (diamonds)  with the dashed 
line showing  the best  fit $<I_{typ}> \sim w^{-1.54}$. See the text for 
details.}  
\label{3}
\end{figure}

Sustitutional disorder introduces in the model a new energy scale that 
competes against the existence of border states carrying anomalously large 
currents. In order to study disorder  effects on the PC it is necessary 
to determine the range of values of disorder $w$ that correspond 
to the diffusive regime in our samples. We begin
analyzing the scaling of $< I_{typ}>$ with $w$ for the SL.
According to elementary scattering theory a scaling as  
$< I_{typ}>   \sim w^{-2}$ is predicted in the diffusive regime \cite{mbsf}. 
In Fig.~\ref{3}(a) we 
show with square symbols the Log-Log plot  of  $<I_{typ}>$ at half filling
as a function of $w$ for the  SL in which the total number of 
sites is $N= 2688$.  
We clearly observe a range of values of $w$ between  1 and 4,
 before  the exponential damping characterizing the onset  of the localized
regime,  in which the scaling is  $ <I_{typ}> \sim w^{- \gamma}$, with
$\gamma = 1.98$ obtained from the best  fit.
Therefore, this  range of $w$ should correspond  to the diffusive regime 
for other geometries containing the same number of total sites. 
Indeed we have obtained in the same range of values of $w$ a similar
 scaling law for a rectangular loop with $N= 2424$
(both rectangles in the loop  having the same aspect ratio), 
even though this loop is not a {\it symmetric} loop 
(in the sense defined above).
Numerical calculations performed on cylindrical geometries satisfied 
the scaling law $ <I_{typ}> \sim w^{- 2}$ \cite{mbsf,cheung3}.

\begin{figure}
\epsfysize=3.5in
\centerline{\epsffile{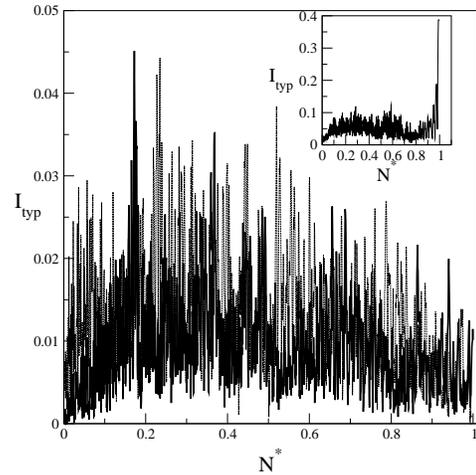}}
\epsfxsize=3.5in
\narrowtext
\caption{$I_{typ}$ as a function of the rescaled number of levels 
$N^{*} \equiv n / N$ for the annular loop AL in the diffusive regime $w =1$ 
(thick solid line) and for the clean pentagon PL (thin dotted line). Both 
curves are almost superimposed. The inset shows for comparison $I_{typ}$ 
as a function of $N^{*}$ for the clean AL.}  
\label{4}
\end{figure}

On the other hand for samples with SR  like the annular loop AL, 
the dependence of  $ <I_{typ}>$ at half filling with $w$ shows a non monotonic
behavior that changes the overall value of the exponent to  $\gamma \neq 2 $ 
(see the circles in Fig.~\ref{3}(a)). 
We argue that by effect of the bulk disorder (and before the onset of the 
localized regime) the states trapped by the SR are unpined being able 
to  carry a finite PC. 
Moreover, there is no reason a priori to expect the same degree of response
to  a  perturbation for all  the trapped states.
This explains  the step-like behavior   of $ <I_{typ}>$ with $w$ observed
 in Fig.~\ref{3}(a) for the AL.
This non-monotonic response of the PC to  disorder translates, in the case
of the annular sample,  to an
average exponent $\gamma \approx 1.3$, significantly lower than in the case 
of the square loop SL. 
As a consequence, as can be observed from Fig.~\ref{3}(a)
there is range of values $w$ where the  PC in 
diffusive annular loops are larger than those in square loops,  
despite the SR present in the former samples. 
This overall  change in the exponent characterizing the diffusive regime 
does not seem to be  a particular attribute of samples with SR. 
For other non-integrable geometries without SR, like rectangular loops where
the inner an outer border have different shapes (chaotic geometry), we also 
find an anomalous scaling  $ <I_{typ}> \sim w^{-\gamma}$ with
$\gamma < 2$. Nevertheless in the absence of SR
we do not observe the non-monotonic response  characteristic of  samples 
with SR. Fig.~\ref{3}(b) shows the Log-Log plot  of 
$ <I_{typ}>$ as a function of $w$ for a non-symmetric rectangular loop 
formed by an inner rectangle of 
sides $l_{1}=13$ and $l_{2}= 8$ and an external rectangle of sides
$L_{1}= 53$ and $L_{2}=48$, respectively. The obtained best  fit  is
 $ <I_{typ}> \sim w^{-  1.54}$.
We stress that in the absence of disorder  this  geometry does not
present border states because the symmetry condition that leads to the peak 
in the DOS is not satisfied. In other words,  $I_{o}$ for a clean 
non-symmetric loop (see Fig.~\ref{4} dotted line) is much smaller than $I_{o}$
for clean {\rm symmetric} loops (see inset of Fig.~\ref{4}).
Even more,  $I_{typ}$   for diffusive {\rm symmetric} loops
can be of the same order of magnitude than $ I_{typ}$ for a 
clean non-symmetric loop, as it is shown in Fig.~\ref{4} for a clean 
pentagonal loop PL  and an annular loop AL with $w=1$. 

In order to determine for a  given diffusive sample  the value of $<I_{typ}>$ 
at half filling, two factors must be considered: 
the value of the current in the clean system ($I_{o}$) and the correction  
factor that takes into account the effect of disorder. 
For generic non-symmetric systems, the exponent $\gamma$ that takes 
into account the diffusive corrections is  $\gamma < 2$
but, as we previously explained, no enhancement of  the value of $I_{o}$ is 
expected. On the other hand, in clean {\rm symmetric loops} without SR  
due to the presence of  border states,  large values of $I_{o}$ can be 
obtained, although in the diffusive regime the exponent is $\gamma= 2$. 
It is for  clean {\rm symmetric loops} with SR, that we find an optimal
interplay  between geometrical effects and bulk disorder. That is
large values of $I_{o}$ compared to those of generic samples, and
an anomalous scaling   $ <I_{typ}> \sim w^{-\gamma}$ with  $\gamma < 2$.
Therefore  these  systems are candidates to have the largest  values of
persistent currents even deep in the diffusive regime.
Also, we obtain that in symmetric loops (with or without SR) the response 
of the system to the external flux is always paramagnetic, even in the 
difussive regime.

The novel effects described in this work can be related to up to date 
unexplained experimental data on PC in mesoscopic metallic rings. 
In Ref.\cite{chandra} rectangular and
circular $Au$ loops with a large number of channels were designed and
the  PC measured. The corresponding signals were 
paramagnetic and more than an order of magnitude larger  than  
predicted values that include the diffusive  correction factor. 
In addition, in that experiment the reported value of the signal for the 
annular loops was $5$ times larger than for
the rectangular loop (although  the estimated theoretical values were 
roughly the same).  
We believe that in order to model the samples employed in this experiment, 
geometries with decoupled degrees of freedom (like the cylinders employed in 
many theoretical papers)  are not appropriate. We present numerical evidence 
showing that by  breaking  the translational  symmetry  in all directions,
it is possible to  create at finite magnetic flux  border states near  
half filling. This states  coherently contribute with  large values of PC. 
As we have previously shown, the confinement in general gives rise to 
geometrical disorder (SR) that competes  with the sustitutional disorder. 
This results in a  non-monotonic behavior of $<I_{typ} >$ with $w$ and
an overall exponent $\gamma < 2$. 
Therefore, we can obtain for  these systems larger values
of PC than in previous models. Moreover, we 
find that the symmetric annular loops can have in the diffusive regime a PC as 
larger as in generic $\rm clean$ systems. 
In addition we show  that  in generic non-integrable geometries without SR the 
exponent $\gamma$ is also $< 2$, although the response of $<I_{typ} >$  with
$w$  is monotonic due to the absence of trapped states in the clean regime. 
This anomalous scaling of $<I_{typ} >$ with $w$  could be a quantum 
signature of the classical non-integrability of the loops. 
The absence of symmetries in the clean system makes 
the effect of the sustitutional disorder not such strong as in the 
case of systems that  in the clean regime preserve some rotational symmetry.   
Indeed, for clean integrable geometries, like the  squares and  symmetric 
rectangular loops presented in this work, and also in  cylinders \cite{mbsf},
the exponent is $\gamma \sim 2$.
We believe that due to  recent technological developments it should be 
feasible to design samples with specific symmetries in which our predictions 
could be confirmed.
 
This work was partially supported by UBACYT (TW35-TX61), CONICET and 
Fundaci\'on Antorchas.

\end{multicols}


\begin{thebibliography}{10}
\bibitem{note} For example, a  square loop is a square  cavity  of side  
$L_1$ with a square hole of side $ L_2 < L_1 $ in its center.
\end{thebibliography}
\end{document}